\documentclass[prd,showpacs,twocolumn,aps]{revtex4}

\usepackage{amsmath}
\usepackage{amssymb}
\usepackage{latexsym}
\usepackage{amsfonts}
\usepackage{epsfig}
\usepackage{psfrag}
\usepackage{graphicx}

\usepackage{color}

\definecolor{black}{rgb}{0,0,0}
\definecolor{blue}{rgb}{0,0,1}
\definecolor{green}{rgb}{0,1,0}
\definecolor{red}{rgb}{1,0,0}
\definecolor{brown}{rgb}{0.4,0.2,0}
\definecolor{darkgreen}{rgb}{0,0.7,0}

\newcommand{\nn}{\nonumber\\}

\newcommand{\f}[1]{\mbox{\boldmath$#1$}}
\newcommand{\fk}[1]{\mbox{\boldmath$\scriptstyle#1$}}

\newcommand{\na}{\mbox{\boldmath$\nabla$}}
\newcommand{\bea}{\begin{eqnarray}}
\newcommand{\ea}{\end{eqnarray}}
\newcommand{\eea}{\end{eqnarray}}
\newcommand{\ord}{\,{\cal O}}

\begin{document}

\title{Momentum dependence in the dynamically assisted Sauter-Schwinger effect}
 
\author{Christian Fey and Ralf Sch\"utzhold}
\email[e-mail:\,]{ralf.schuetzhold@uni-due.de}

\affiliation{Fakult\"at f\"ur Physik, Universit\"at Duisburg-Essen, 
Lotharstrasse 1, 47057 Duisburg, Germany}
 
\date{\today}

\begin{abstract}
Recently it has been found that the superposition of a strong and slow 
electric field with a weaker and faster pulse can significantly enhance 
the probability for non-perturbative electron-positron pair creation out 
of the vacuum -- the dynamically assisted Sauter-Schwinger effect.
Via the WKB method, we estimate the momentum dependence of the pair 
creation probability and compare it to existing numerical results. 
Besides the theoretical interest, a better understanding of this 
pair creation mechanism should be helpful for the planned experiments 
aiming at its detection. 
\end{abstract}


\maketitle

%%%%%%%%%%%%%%%%%%%%%%%%%%%%%%%%%%%%%%%%%%%%%%%%%%%%%%%%%%%%%%%%%%%%%%%%%%%%%
\section{Introduction}
%%%%%%%%%%%%%%%%%%%%%%%%%%%%%%%%%%%%%%%%%%%%%%%%%%%%%%%%%%%%%%%%%%%%%%%%%%%%%

Quantum electrodynamics (QED), the theory of charged particles such as 
electrons and positrons interacting with the electromagnetic field, 
offers many interesting phenomena beyond standard perturbation theory 
(i.e., expansion into powers of the fine-structure constant 
$\alpha_{\rm QED}$). 
One prominent example is the Sauter-Schwinger effect 
\cite{Sauter,Euler,Schwinger}
describing the creation of electron-positron pairs out of the QED vacuum 
by an electric field via tunneling.
For a constant electric field $E$, the lowest-order pair creation probability 
scales as 
\bea
\label{probability}
P_{e^+e^-}
\sim 
\exp\left\{-\pi\,\frac{m^2}{qE}\,\frac{c^3}{\hbar}\right\}
=
\exp\left\{-\pi\,\frac{E_S}{E}\right\}
\,,
\ea
where $m$ is the mass and $\pm q$ the charge of the electrons and positrons.
Since this expression does not permit a Taylor expansion in $q$,
this phenomenon is a non-perturbative effect.
Because the characteristic field strength $E_S$ 
(which is also called the critical field) 
is very large $\sim10^{18}\rm V/m$,
this fundamental QED prediction has not been conclusively 
experimentally verified yet 
-- in contrast to electron-positron pair creation in the perturbative 
multi-photon regime, see, e.g., \cite{SLAC}. 

Furthermore, in spite of many efforts, see, e.g., 
\cite{Nikishov+Ritus,Brezin+Itzykson,Dumlu,Keitel,Hebenstreit,diverse},
our understanding of this non-perturbative effect for non-constant 
electric fields $E$ is still far from complete.
For example, recently it has been found \cite{Assisted} that the 
superposition of a strong and slow electric field with a weaker and 
faster pulse can significantly enhance the pair creation probability,
see also \cite{Monin+Voloshin,Catalysis,Orthaber}.
In the following, we are going to present an analytic estimate of the 
momentum dependence of this enhancement mechanism via the WKB  
approximation (for more details, see also \cite{Fey}).
These findings should be relevant for the planned experiments with 
optical lasers or XFEL (or a combination of both), see, e.g., 
\cite{Ringwald}, which might facilitate the first conclusive
observation of this effect.

%%%%%%%%%%%%%%%%%%%%%%%%%%%%%%%%%%%%%%%%%%%%%%%%%%%%%%%%%%%%%%%%%%%%%%%%%%%%%
\section{Basic formalism}
%%%%%%%%%%%%%%%%%%%%%%%%%%%%%%%%%%%%%%%%%%%%%%%%%%%%%%%%%%%%%%%%%%%%%%%%%%%%%

We consider quantized electrons and positrons in an external (classical) 
electric field $\f{E}$ with no magnetic field present $\f{B}=0$.
The field $E$ is supposed to be purely time-dependent $\f{E}=E(t)\f{e}_x$ 
and sub-critical $E\ll E_S$.  
Thus we can neglect the spin of the electrons and positrons and use the 
Klein-Fock-Gordon equation in temporal gage $\f{E}=\f{\dot A}$ 
where $\f{A}$ is the vector potential ($\hbar=c=1$)
\bea
\label{Klein-Fock-Gordon}
\left(\frac{\partial^2}{\partial t^2}-
\left[\na-iq\f{A}(t)\right]^2+m^2\right)\Phi=0
\,.
\ea
After a spatial Fourier transform, we get \cite{momentum} 
\bea
\label{scattering}
\left(
\frac{d^2}{dt^2}
+[k_x-qA(t)]^2
+\f{k}_\perp^2
+m^2
\right)
\phi_{\fk{k}}
=0
\,.
\ea
We see that the momentum $\f{k}_\perp$ transversal to the electric field
can be absorbed by a re-definition of the effective mass 
$m^2 \to m_{\rm eff}^2=\f{k}_\perp^2+m^2$.
Thus, we omit it in the following {formul\ae} for brevity. 

The above equation is formally equivalent to a harmonic oscillator with 
a time-dependent potential
\bea
\label{oscillator}
\left(\frac{d^2}{dt^2}+\Omega^2(t)\right)\phi(t)=0
\,.
\ea
If we replace $t$ by $x$, $\phi(t)$ by $\psi(x)$, 
and $\Omega^2(t)$ by $2m[E-V(x)]$, 
this equation has the same form as a one-dimensional 
Schr\"odinger scattering problem with energy $E$
and potential $V(x)$.  
Therefore, a solution which initially behaves as 
$\phi_{\fk{k}}^{\rm in}(t)=\exp\{-i\omega_{\fk{k}}t\}$
will finally evolve into a mixture of positive and negative 
frequencies 
\bea
\label{mixing}
\phi_{\fk{k}}^{\rm out}(t)
=
\alpha_{\fk{k}}\exp\{-i\omega_{\fk{k}}t\}+
\beta_{\fk{k}}\exp\{+i\omega_{\fk{k}}t\}
\,,
\ea
where the Bogoliubov coefficients $\alpha_{\fk{k}}$ and 
$\beta_{\fk{k}}$ are related to the reflection $R$ and 
transmission $T$ amplitudes in the one-dimensional 
scattering theory picture via $\alpha_{\fk{k}}=1/T$
and $\beta_{\fk{k}}=R/T$. 
The probability for electron-positron pair creation 
is given by $P_{e^+e^-}=\sum_{\fk{k}}|\beta_{\fk{k}}|^2$.

%%%%%%%%%%%%%%%%%%%%%%%%%%%%%%%%%%%%%%%%%%%%%%%%%%%%%%%%%%%%%%%%%%%%%%%%%%%%%
\section{Riccati equation}
%%%%%%%%%%%%%%%%%%%%%%%%%%%%%%%%%%%%%%%%%%%%%%%%%%%%%%%%%%%%%%%%%%%%%%%%%%%%%

For slowly varying and sub-critical electric fields $E(t)$, 
the Bogoliubov coefficients $\alpha_{\fk{k}}$ and $\beta_{\fk{k}}$
can be derived via the WKB approximation.
To this end, let us cast Eq.~(\ref{oscillator}) into a first-order form 
\bea
\label{first-order}
\frac{d}{dt}
\left(
\begin{array}{c}
\phi \\
\dot\phi
\end{array}
\right)
=
\f{\dot v}
=
\left(
\begin{array}{cc}
0 & 1 \\
-\Omega^2(t) & 0
\end{array}
\right)
\cdot
\left(
\begin{array}{c}
\phi \\
\dot\phi
\end{array}
\right)
=
\f{M}\cdot\f{v}
\,.
\ea
Since usual scalar product is not conserved by this evolution equation, 
it is useful to introduce the inner product 
\bea
\label{inner}
(\f{v}|\f{v'})=-i(v_1^*v_2'-v_2^*v_1')
\,,
\ea
which is just the Wronskian of the original Eq.~(\ref{oscillator}).
Thus, the inner product of two solutions $\f{v}$ and $\f{v'}$ 
is conserved 
\bea
\label{conserved}
\frac{d}{dt}(\f{v}|\f{v'})=0
\,.
\ea
As the next step, we expand the solution $\f{v}(t)$ of 
Eq.~(\ref{first-order})
\bea
\label{expand}
\f{v}(t)
=
\alpha(t)e^{i\varphi(t)}\f{v}_+(t)
+
\beta(t)e^{-i\varphi(t)}\f{v}_-(t)
\,,
\ea
into instantaneous eigenvectors $\f{v}_\pm(t)$ of the matrix 
\bea
\label{eigenvectors}
\f{M}\cdot\f{v}_\pm(t)
=
\pm i\Omega(t)\f{v}_\pm(t)
\,.
\ea
With the normalization $\f{v}_\pm=(1,\pm i\Omega)^T/\sqrt{2\Omega}$, 
we find 
$(\f{v}_+|\f{v}_+)=1$, $(\f{v}_-|\f{v}_-)=-1$, and $(\f{v}_+|\f{v}_-)=0$.
Finally, $\alpha(t)$ and $\beta(t)$ are the instantaneous
Bogoliubov coefficients, where we have separated out the WKB phase 
\bea
\label{phase}
\varphi(t)=\int\limits_{-\infty}^t dt'\,\Omega(t')
\,.
\ea
This uniquely determines the evolution of $\alpha(t)$ and $\beta(t)$
which we can obtain by inserting the expansion (\ref{expand}) into
the equation of motion (\ref{first-order}) and projecting it with the 
inner product (\ref{inner}) onto the eigenvectors $\f{v}_\pm(t)$  
\bea
\label{uniquely}
\dot\alpha(t)=\frac{\dot\Omega(t)}{2\Omega(t)}\,e^{-2i\varphi(t)}\beta(t)
\,,\;
\dot\beta(t)=\frac{\dot\Omega(t)}{2\Omega(t)}\,e^{2i\varphi(t)}\alpha(t)
\,,
\ea
where we have used 
$(\f{v}_+|\f{\dot v}_+)=(\f{v}_-|\f{\dot v}_-)=0$, as well as 
$(\f{v}_-|\f{\dot v}_+)=\dot\Omega/(2\Omega)$ and 
$(\f{v}_+|\f{\dot v}_-)=-\dot\Omega/(2\Omega)$. 
In terms of the reflection coefficient $R(t)=\beta(t)/\alpha(t)$, we get 
\bea
\label{Riccati}
\dot R(t)=\frac{\dot\Omega(t)}{2\Omega(t)}
\left(e^{2i\varphi(t)}-R^2(t)e^{-2i\varphi(t)}\right)
\,,
\ea
which is known as Riccati equation, see also \cite{Dumlu,Riccati}.

%%%%%%%%%%%%%%%%%%%%%%%%%%%%%%%%%%%%%%%%%%%%%%%%%%%%%%%%%%%%%%%%%%%%%%%%%%%%%
\section{WKB method}%\paragraph*{WKB method}
%%%%%%%%%%%%%%%%%%%%%%%%%%%%%%%%%%%%%%%%%%%%%%%%%%%%%%%%%%%%%%%%%%%%%%%%%%%%%

The Riccati equation (\ref{Riccati}) is still exact but unfortunately 
non-linear.
The WKB approximation is based on the assumption that the rate of 
change of $\Omega(t)$ is much slower than the internal frequency 
$\Omega(t)$ itself.
In our case, 
\bea
\Omega^2(t)
=
[k_x-qA(t)]^2+m^2
\,,
\ea
this is satisfied if the strength $E$ and the rate of change $\omega$
of the electric field $E(t)$ are small compared to the mass, i.e.,
for $E\ll E_S$ and $\omega\ll m$. 
In this limit, the phase factors $e^{\pm2i\varphi}$ are rapidly oscillating 
and the magnitude of $R$ can be estimated by analytic continuation to 
the complex plane.
In terms of the phase variable $\varphi$, the  Riccati equation 
(\ref{Riccati}) reads 
\bea
\frac{dR(\varphi)}{d\varphi}=\frac12
\left(e^{2i\varphi}-R^2(\varphi)e^{-2i\varphi}\right)
\frac{d\ln\Omega(\varphi)}{d\varphi}
\,.
\ea
Now analytic continuation to the upper complex half-plane 
$\varphi\to\Re\varphi+i\Im\varphi=\Re\varphi+i\chi$ 
shows that $R$ becomes exponentially suppressed $R\sim e^{-2\chi}$.
The degree of this suppression depends on the point where the 
analytic continuation breaks down.
Since $e^{\pm2i\varphi}$ is analytic everywhere, this will be 
determined by the term $\ln\Omega$.
Typically, one can go into the upper complex half-plane until one 
hits the first zero of $\Omega$ at $t_*$, i.e., $\Omega(t_*)=0$. 
These points $t_*$ in the complex plane are analogous to the 
classical turning points in WKB.
Consequently, we find 
\bea
R(t\uparrow\infty)=
R\sim e^{-2\chi_*}
=
\exp\left\{-2\Im\left[\varphi(t_*)\right]\right\}
\,.
\ea
So far, this is only an order-of-magnitude estimate.
However, it can be shown \cite{Davis} that this expression becomes exact 
(under appropriate conditions) in the adiabatic limit 
(roughly speaking, $m\uparrow\infty$), i.e., that the pre-factor in 
front of the exponential tends to one. 

In case of more than one turning point, the one with the 
smallest $\chi_*$, i.e., closest to real axis 
(in the complex $\varphi$-plane) dominates.
For multiple turning points with similar $\chi_*$, 
there can be interference effects \cite{Dumlu,Hebenstreit}.  

%%%%%%%%%%%%%%%%%%%%%%%%%%%%%%%%%%%%%%%%%%%%%%%%%%%%%%%%%%%%%%%%%%%%%%%%%%%%%
\section{Double Sauter pulse}
%%%%%%%%%%%%%%%%%%%%%%%%%%%%%%%%%%%%%%%%%%%%%%%%%%%%%%%%%%%%%%%%%%%%%%%%%%%%%

Now we are in the position to apply the above method to the double 
Sauter pulse studied in \cite{Assisted}
\bea
\label{double-E}
E(t)=\frac{E_1}{\cosh^2(\omega_1t)}+\frac{E_2}{\cosh^2(\omega_2t)}
\,.
\ea
Note that the Dirac equation in the presence of a single pulse 
(e.g., $E_2=0$) of that type can be solved exactly.
This solution has already been used by Sauter \cite{Sauter} even though 
with $t$ and $x$ interchanged. 

The first term on the r.h.s.\ describes a strong and slow electric
field profile while the second term corresponds to a much weaker and 
faster pulse with $\omega_1\ll\omega_2\ll m$ and  
$E_2 \ll E_1 \ll E_S$. 
The Keldysh \cite{Keldysh} parameters of the two pulses are supposed to be 
small $\gamma_1=m\omega_1/(qE_1)\ll1$ and large 
$\gamma_2=m\omega_2/(qE_2)\gg1$, respectively, while the 
combined Keldysh parameter introduced in \cite{Assisted}
\bea
\label{combined}
\gamma_{\rm c}=\frac{m\omega_2}{qE_1}
\,,
\ea
is of order one. 
For time scales where the second (fast) pulse contributes, 
we may approximate the first (slow) pulse by a constant field 
$E_1$ due to $\omega_1\ll\omega_2$. 
Thus the vector potential becomes 
\bea
\label{double-A}
A(t)\approx E_1t
+\frac{E_2}{\omega_2}\,\tanh(\omega_2t)
\,, 
\ea
and the condition $\Omega(t_*)=0$ for the turning points reads 
\bea
\label{turning}
\omega_2t_*+\frac{E_2}{E_1}\,\tanh(\omega_2t_*)
=
\gamma_{\rm c}
\left(i+\frac{k_x}{m}\right)
\,.
\ea
Due to the periodicity of the function $\tanh(\omega_2t)$,
there are infinitely many solutions.
However, we are mainly interested in those close to the real 
axis -- which have the smallest value of $\chi_*$.
For $E_1\gg E_2$, we may approximate the two most relevant
solutions $t_*$ and $t_*'$ via 
\bea
\label{relevant}
t_*\approx
\frac{m}{qE_1}
\left(i+\frac{k_x}{m}\right)
\,,\quad 
t_*'\approx
\frac{i\pi}{2\omega_2}
\,.
\ea
Note that this approximation breaks down when $\gamma_{\rm c}$
approaches $\pi/2$, so we will assume $\gamma_{\rm c}>\pi/2$
in the following, see also \cite{Assisted}. 
The first solution $t_*$ is basically the same as for the slow 
pulse alone (i.e., $E_2=0$), while the second 
one $t_*'$ is obviously tied to the fast pulse. 

Again using $E_1\gg E_2$, we may derive the associated 
exponents $\chi_*$ and $\chi_*'$.
For the normal solution $t_*$, we reproduce the usual value 
given solely by the slow pulse 
\bea
\label{normal}
\chi_*=\frac{\pi m^2}{4qE_1}
\,.
\ea
With $P_{e^+e^-}\sim|\beta_{\fk{k}}^2|\approx|R^2|\sim e^{-4\chi_*}$, 
we verify Eq.~(\ref{probability}).  
This result is independent of $k_x$ because we 
have approximated the slow pulse by a constant field. 
The anomalous solution $t_*'$, on the other hand, yields 
\bea
\label{anomalous}
\chi_*'
=
\frac{m^2}{2qE_1}
\,\Im\left\{ 
f\left(\frac{i\pi}{2\gamma_{\rm c}}-\frac{k_x}{m}\right)
\right\}
\,,
\ea
with $f(x)=x\sqrt{1+x^2}+{\rm arcsinh}(x)$. 
In this form, the dependence on $k_x$ is perhaps not so obvious.
Thus, let us Taylor expand this expression in powers of $k_x$
\bea
\label{anomalous-Taylor}
\chi_*'
&=&
\frac{m^2}{2qE_1}
\left[
\frac{\pi}{2\gamma_{\rm c}}
\sqrt{1-\left(\frac{\pi}{2\gamma_{\rm c}}\right)^2}
+
{\rm arcsin}\left(\frac{\pi}{2\gamma_{\rm c}}\right)
\right]
+
\nn
&&
+\frac{k_x^2}{2qE_1}\,\frac{\pi}{\sqrt{4\gamma_{\rm c}^2-\pi^2}}
+\ord(k_x^4)
\,.
\ea
The momentum-independent result in the first line was already derived 
in \cite{Assisted} via the word-line instanton method, but this method 
did not yield any information about the momentum dependence in the second 
line.

%%%%%%%%%%%%%%%%%%%%%%%%%%%%%%%%%%%%%%%%%%%%%%%%%%%%%%%%%%%%%%%%%%%%%%%%%%%%%
\section{Discussion}
%%%%%%%%%%%%%%%%%%%%%%%%%%%%%%%%%%%%%%%%%%%%%%%%%%%%%%%%%%%%%%%%%%%%%%%%%%%%%

%%%%%%%%%%%%%%%%%%%%%%%%%%%%%%%%%%%%%%%%%%%%%%%%%%%%%%%%%%%%%%%%%%%%%%%%%%%%%
\begin{figure}[h]
\includegraphics[width=.9\columnwidth]{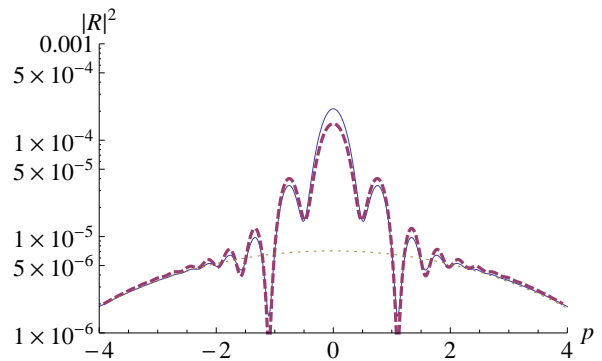}
\caption{(Color online) Comparison of analytical and numerical 
results for the (differential) pair creation probability $|R|^2$ depending 
on the momentum $p=k_x/m_{\rm eff}$. 
The solid (blue) curve shows the numerical data from \cite{Orthaber}
(kindly provided by the authors)  for $E_1=E_S/4$, $\omega_1=10^4\rm eV$, 
$E_2=E_1/10$, and $\omega_2=5\times10^5\rm eV$.
For the same values, the function 
$|R|^2=|c_*\exp\{-2\varphi(t_*)\}+c_*'\exp\{-2\varphi(t_*')\}|^2$ 
representing the WKB result is plotted (dashed purple curve), 
where $t_*$ and $t_*'$ are the (exact) WKB turning points 
and $\varphi$ is the complex phase in Eq.~(\ref{phase}). 
The pre-factors are chosen as $c_*=1.4$ and $c_*'=0.2$. 
Finally, the dotted (brown) curve describes the case $E_2=0$ of a 
single pulse (i.e., the background).}
\label{figure}
\end{figure}
%%%%%%%%%%%%%%%%%%%%%%%%%%%%%%%%%%%%%%%%%%%%%%%%%%%%%%%%%%%%%%%%%%%%%%%%%%%%%

For the dynamically assisted Sauter-Schwinger effect \cite{Assisted},
we estimated the momentum dependence in Eqs.~(\ref{anomalous}) and 
(\ref{anomalous-Taylor}) via the WKB method. 
The strong and slow pulse alone is represented by the normal turning 
point $t_*$ with $\chi_*$ in Eq.~(\ref{normal}).
It has a broad momentum distribution $(\Delta k_x)^2/m^2\sim1/\gamma_1^2\gg1$ 
which can be explained by the uncertainty $\Delta t\sim1/\omega_1$ 
of the creation time: 
Particles that are created earlier have more time to be accelerated 
by the strong electric field than others which are produced later
\cite{momentum}. 
The dynamically assisted effect, on the other hand, corresponds to 
the anomalous turning point $t_*'$ determined by the weak and fast pulse.
Thus, the creation time is far less uncertain $\Delta t\sim1/\omega_2$
and -- as one would expect -- it has a much narrower width, 
even though it is probably hard to guess the precise scaling  
$(\Delta k_x)^2\sim qE\sqrt{4\gamma_{\rm c}^2-\pi^2}$ beforehand.  
Consistent with \cite{Assisted}, this window closes for 
$\gamma_{\rm c}\downarrow\pi/2$, after which the enhancement 
disappears. 
For $\gamma_{\rm c}<\pi/2$, the normal turning point $t_*$ dominates
$\chi_*<\chi_*'$. 

Let us compare our findings with the numerical results in \cite{Orthaber}
obtained via the quantum kinetic approach. 
In Fig.~\ref{figure}, we plot the numerical data from Fig.~5a of 
\cite{Orthaber} as well as  
$|R|^2=|c_*\exp\{-2\varphi(t_*)\}+c_*'\exp\{-2\varphi(t_*')\}|^2$, 
where $\varphi$ is the complex phase in Eq.~(\ref{phase}). 
The two WKB turning points $t_*$ and $t_*'$ can be obtained exactly, 
but our WKB method does not yield the pre-factors $c_*$ and $c_*'$,
which can also depend on $p$ in general. 
Neglecting this $p$-dependence, we may estimate the normal pre-factor 
$c_*$ by comparison to the case of a single pulse (i.e., $E_2=0$) which 
facilitates an analytic solution.  
The anomalous pre-factor $c_*'$, on the other hand, was chosen (fitted) 
to match the numerical results. 
After that, the agreement between the analytic and the numerical results 
is surprisingly good -- given that the employed values 
$E_1=E_S/4$, $\omega_1=10^4\rm eV$, $E_2=E_1/10$, and 
$\omega_2=5\times10^5\rm eV$ do not satisfy our underlying 
assumptions $E_2 \ll E_1 \ll E_S$ and $\omega_1\ll\omega_2\ll m$
very well. 
Note that the difference between $c_*$ and $c_*'$ also indicates that 
we are not deep in the adiabatic limit ($m\uparrow\infty$). 
Furthermore, one should be very careful with the order of the various 
limits in this multiple-scale problem. 
For example, the adiabatic limit ($m\uparrow\infty$) does not commute 
with the limit $E_2/E_1\downarrow0$ since the dynamically assisted 
Sauter-Schwinger effect given by $c_*'$ should vanish for $E_2=0$. 
Finally, the oscillations visible in Fig.~\ref{figure} 
(and Fig.~5a of \cite{Orthaber}) can be explained nicely by interference 
effects \cite{Dumlu,Hebenstreit} of the two turning points $t_*$ and $t_*'$.
The interferences are most pronounced where the two contributions 
$c_*\exp\{-2\varphi(t_*)\}$ and $c_*'\exp\{-2\varphi(t_*')\}$ are 
equally strong, which happens around $k_x\approx\pm m_{\rm eff}$
in this case.

%%%%%%%%%%%%%%%%%%%%%%%%%%%%%%%%%%%%%%%%%%%%%%%%%%%%%%%%%%%%%%%%%%%%%%%%%%%%%
\section{Outlook}
%%%%%%%%%%%%%%%%%%%%%%%%%%%%%%%%%%%%%%%%%%%%%%%%%%%%%%%%%%%%%%%%%%%%%%%%%%%%%

It might be interesting to generalize the above findings to other 
pulse profiles such as 
\bea
E(t)=E_1f_1(\omega_1t)+E_2f_2(\omega_2t)
\,,
\ea
with $f_1(0)=f_2(0)=1$ as well as $\omega_1\ll\omega_2\ll m$ and  
$E_2 \ll E_1 \ll E_S$. 
For a broad class of functions, for example $f_2(x)=1/(1+x^2)^2$,
we expect to obtain qualitatively the same picture as discussed above.
Again, the normal turning point $t_*$ will have basically the same 
value as in Eq.~(\ref{relevant}) and 
the anomalous turning point $t_*'$ will be very close to the 
singularity of $f_2$, in our example $t_*'\approx i/\omega_2$.  
As before, for small values of the combined Keldysh parameter
$\gamma_{\rm c}$ in Eq.~(\ref{combined}), the contribution 
of the normal turning point $t_*$ dominates -- 
but if $\gamma_{\rm c}$ exceeds a critical value of order one, 
the anomalous turning point $t_*'$ becomes stronger than the normal 
one and we get dynamically assisted pair creation.

For other profiles, such as $f_2(x)=\exp\{-x^2\}$, however, the anomalous 
turning point $t_*'$ logarithmically depends on the ratio $E_1/E_2$ and 
thus the mechanism of dynamically assisted pair creation 
(including the threshold value for $\gamma_{\rm c}$) 
will also depend on $E_1/E_2$, in contrast to the case considered above.

%%%%%%%%%%%%%%%%%%%%%%%%%%%%%%%%%%%%%%%%%%%%%%%%%%%%%%%%%%%%%%%%%%%%%%%%%%%%%
\begin{acknowledgments}
We thank the authors of \cite{Orthaber} for kindly providing 
their numerical data.
R.S.\ gratefully acknowledges fruitful discussions with 
G.~Dunne \& H.~Gies and financial support by the DFG. 
\end{acknowledgments}
%%%%%%%%%%%%%%%%%%%%%%%%%%%%%%%%%%%%%%%%%%%%%%%%%%%%%%%%%%%%%%%%%%%%%%%%%%%%%

\bibliographystyle{apsrmp4}

\begin{thebibliography}{99}
\bibitem{momentum}
Since Eq.~(\ref{Klein-Fock-Gordon}) is spatially homogeneous, 
the canonical momentum $\f{k}$ is conserved -- but the physical momentum 
$\f{p}_{\rm phys}=\f{k}\pm q\f{A}$ is not 
(as the particles are accelerated by the electric field).
\bibitem{Riccati}
Strictly speaking, Eq.~(\ref{Riccati}) is the Riccati equation for scalar 
QED since we started from the Klein-Fock-Gordon equation 
(\ref{Klein-Fock-Gordon}) and thus neglected the spin of the electrons 
and positrons. 
%
Taking their spin into account, the Riccati equation for spinor QED has 
the same form as Eq.~(\ref{Riccati}) after replacing $\dot\Omega$ by 
another time-dependent function, see, e.g., \cite{Dumlu}. 
%
However, since the exponential dependence $e^{\pm2i\varphi}$ and the 
location $t_*$ of the singularities $\Omega(t_*)=0$ is the same in both 
cases, the WKB exponents $e^{-2\chi_*}$ calculated here remain unchanged, 
only the pre-factors (including relative phases) may be different.
\end{thebibliography}

%%%%%%%%%%%%%%%%%%%%%%%%%%%%%%%%%%%%%%%%%%%%%%%%%%%%%%%%%%%%%%%%%%%%%%%%%%%%%

%%%%%%%%%%%%%%%%%%%%%%%%%%%%%%%%%%%%%%%%%%%%%%%%%%%%%%%%%%%%%%%%%%%%%%%%%%%%%
%%%%%%%%%%%%%%%%%%%%%%%%%%%%%%%%%%%%%%%%%%%%%%%%%%%%%%%%%%%%%%%%%%%%%%%%%%%%%
%%%%%%%%%%%%%%%%%%%%%%%%%%%%%%%%%%%%%%%%%%%%%%%%%%%%%%%%%%%%%%%%%%%%%%%%%%%%%
%%%%%%%%%%%%%%%%%%%%%%%%%%%%%%%%%%%%%%%%%%%%%%%%%%%%%%%%%%%%%%%%%%%%%%%%%%%%%
%%%%%%%%%%%%%%%%%%%%%%%%%%%%%%%%%%%%%%%%%%%%%%%%%%%%%%%%%%%%%%%%%%%%%%%%%%%%%
%%%%%%%%%%%%%%%%%%%%%%%%%%%%%%%%%%%%%%%%%%%%%%%%%%%%%%%%%%%%%%%%%%%%%%%%%%%%%
\end{document}